\documentclass[11pt,fleqn,twoside]{article}
\usepackage{amsfonts,amssymb,latexsym}
\makeatletter
\newcommand{\prava}[1]{\small\it
\begin{flushleft}
Copyright \copyright \ 1999 by  #1
\end{flushleft}}

\newcommand{\name}[1]{\begin{flushleft}
                       \LARGE \bf #1
                       \end{flushleft}\vspace{-3mm}}

\newcommand{\Author}[1]{\begin{flushleft}
                       \it #1 \end{flushleft}}

\newcommand{\Adress}[1]{\begin{flushleft}
                       \it #1 \end{flushleft}}

\newcommand{\Date}[1]{\begin{flushleft}
                      \small  \it #1 \end{flushleft}}

\newcommand{\ehkol}{Author \ name}
\newcommand{\ohkol}{Article \ name}
\renewcommand{\@evenhead}{
\hspace*{-3pt}\raisebox{-15pt}[\headheight][0pt]{\vbox{\hbox to \textwidth 
{\thepage \hfil \ehkol}\vskip4pt \hrule}}}
\renewcommand{\@oddhead}{
\hspace*{-3pt}\raisebox{-15pt}[\headheight][0pt]{\vbox{\hbox to \textwidth 
{\ohkol \hfil \thepage}\vskip4pt\hrule}}}
\renewcommand{\@evenfoot}{}
\renewcommand{\@oddfoot}{}

\newcommand{\be}{\begin{equation}}
\newcommand{\ee}{\end{equation}}
\newcommand{\ba}{\hspace*{-5pt}\begin{array}}
\newcommand{\ea}{\end{array}}

\newcommand{\ds}{\displaystyle}
\makeatother

\newtheorem{thm}{Theorem}[section]
\newtheorem{prop}[thm]{Proposition}
\newtheorem{lem}[thm]{Lemma}

\newcommand{\bra}{_{\epsilon}
\langle \Psi_N(\lambda_1,\ldots,\lambda_N)\vert}
\newcommand{\ket}{\vert \Psi_N(\lambda_1,\ldots,\lambda_N)
\rangle_{\epsilon}}

\begin{document}

\def\theequation{\arabic{section}.\arabic{equation}}
\setcounter{page}{99}
\thispagestyle{empty}

\renewcommand{\ehkol}{T. Kojima}
\renewcommand{\ohkol}{Dynamical Correlation Functions for an
Impenetrable Bose Gas}

\begin{flushleft}
\footnotesize \sf
Journal of Nonlinear Mathematical Physics \qquad 1999, V.6, N~1,
\pageref{kojima-fp}--\pageref{kojima-lp}.
\hfill {\sc Article}
\end{flushleft}

\vspace{-5mm}

\renewcommand{\footnoterule}{}
{\renewcommand{\thefootnote}{} \footnote{\prava{T. Kojima}}}

\name{Dynamical Correlation Functions for \\
an Impenetrable Bose Gas with Neumann \\
or Dirichlet Boundary Conditions}\label{kojima-fp}

\Author{Takeo KOJIMA}

\Adress{Department of Mathematics, College of Science and Technology,
Nihon University,\\
 1-8, Kanda-Surugadai, Chiyoda Tokyo 101, Japan\\[1mm]
E-mail: kojima@math.cst.nihon-u.ac.jp}

\Date{Received September 09, 1998; Accepted September 28, 1998}

\begin{abstract}
\noindent
We study the time and temperature dependent correlation functions
for an impenetrable Bose gas with Neumann or Dirichlet
boundary conditions 
$\langle \psi(x_1,0)\psi^\dagger(x_2,t)\rangle
_{\pm,T}$.
We derive the Fredholm determinant formulae for the correlation
functions, by means of the Bethe Ansatz.
For the special case $x_1=0$,
we express correlation functions with Neumann boundary
conditions 
$\langle\psi(0,0)\psi^\dagger(x_2,t)\rangle
_{+,T}$, in terms of
solutions of nonlinear partial dif\/ferential equations
which were introduced in \cite{kojima:Sl} as a generalization of
the nonlinear Schr\"odinger equations.
We generalize the Fredholm minor determinant formulae
of ground state correlation functions 
$\langle\psi(x_1)\psi^\dagger(x_2)\rangle
_{\pm,0}$ in~\cite{kojima:K},
to the Fredholm determinant formulae for
the time and temperature dependent 
correlation functions
$\langle\psi(x_1,0)\psi^\dagger(x_2,t)\rangle
_{\pm,T}$, $t \in {\bf R}$, $T \geq 0$.
\end{abstract}

\section{Introduction}
In the standard treatment of quantum integrable models,
one starts with 
a f\/inite box and impose periodic boundary conditions,
in order to ensure integrability.
Recently, there has been increasing interest in exploring
other possible boundary conditions compatible with integrability.
These other possible boundary conditions are called 
``open boundary conditions''.

With open boundary conditions, the works on the two dimensional 
Ising model are among the earliest.
By the help of graph theoretical approach, 
B.M.~McCoy and T.T.~Wu \cite{kojima:M.W.}
studied the two dimensional Ising model 
with open boundary conditions.
They calculated the local magnetizations.
E.K.~Sklyanin~\cite{kojima:S}
began the Bethe Ansatz approach to open boundary problems.
M. Jimbo et.al.~\cite{kojima:J} studied
the antiferromagnetic XXZ chains 
with open boundary conditions
and derived an integrable representation
of correlation functions,
using Sklyanin's algebraic Bethe Ansatz framework and
representation theoretical approach invented by Kyoto
school~\cite{kojima:D.F.J.M.N.,kojima:J.M.}. T.~Kojima~\cite{kojima:K} studied
the ground state correlation functions for
an impenetrable Bose gas with open boundary conditions:
\[
\langle \psi(x_1)\psi^\dagger(x_2)\rangle.
\]
Kojima derived the Fredholm minor determinant representations for 
the ground state correlation functions by the help of fermions,
which have the integral kernel:
\[
\frac{\sin(\lambda-\mu)}{\lambda-\mu}\pm
\frac{\sin(\lambda+\mu)}{\lambda+\mu}.\label{kojima:kernel:deg}
\]
The integral intervals depend on the space parameter $x_1$, $x_2$.
In this paper we study an impenetrable Bose gas 
with open boundary conditions.
We are interested in the f\/inite-temperature 
dynamical correlation functions:
\[
\langle \psi(x_1,0)\psi^\dagger(x_2,t)\rangle_{\pm,T}.
\]
We derive the Fredholm determinant representations for
the dynamical correlation functions 
by the coordinate Bethe Ansatz,
which have the integral kernel:
\[
\sqrt{\vartheta(\lambda)}
\left(
L(\lambda,\mu)\pm
L(\lambda,-\mu)\right)
\sqrt{\vartheta(\mu)},
\]
where we have used
\be \label{kojima:kernel-L}
\ba{l}
\ds L(\lambda,\mu)=\frac{e^{-\frac{1}{2}it(\lambda^2+\mu^2)}}
{\lambda-\mu}\left\{
e^{it\lambda^2}\sin(x_1(\lambda-\mu))+
e^{it\mu^2}\sin(x_2(\lambda-\mu))\right.
\vspace{3mm}\\
\qquad \ds \left.
+\frac{2}{\pi}~{\rm P.V.}\int_{-\infty}^{\infty}
\left(\frac{1}{s-\mu}-\frac{1}{s-\lambda}\right)e^{its^2}
\sin((s-\mu)x_1)\sin((s-\lambda)x_2)ds\right\}.
\ea
\ee
Here the notation ${\rm P.V.}$ represents
Cauchy's principle value
and the measure $\vartheta(\lambda)$ is given~by
\be
\vartheta(\lambda)=\frac{1}{1+\exp\left(
\frac{\lambda^2-h}{T}\right)}.\label{kojima:def:Fermi}
\ee
For an impenetrable Bose gas without boundaries,
V.~Korepin and N.~Slavnov~\cite{kojima:K.S.}
has derived the Fredholm determinant formulae for
the dynamical correlation functions.

To describe the $2n$ point dynamical correlation functions
for an impenetrable Bose gas without boundaries,
N.~Slavnov~\cite{kojima:Sl} introduced a system of nonlinear
partial dif\/ferential equations,
which becomes the nonlinear Schr\"odinger equation
in the simplest case.
The generalization of the nonlinear Schr\"odinger equations
has $2n$ time variables $t_j$, $(1\leq j \leq 2n)$
and $2n$ space variables $x_j$, $(1\leq j \leq 2n)$.
In this paper, we consider
the dynamical correlation functions with Neumann
boundary conditions for
the special case that one space parameter 
has the value $x_1=0$:
$\langle \psi(0,0)\psi^\dagger(x,t)\rangle_{T,+}$.
We express the dynamical correlation functions
in terms of a solution of Slavnov's generalization of 
the nonlinear Schr\"odinger equations.
The dif\/ferential equations,
which describe four-point correlation
functions without boundaries:
\[
\langle \psi(x_1,t_1)\psi^\dagger(x_2,t_2)
\psi(x_3,t_3)\psi^\dagger(x_4,t_4)
\rangle_{T},
\]
describe the dynamical correlation functions
with Neumann boundary conditions:
\[
\langle \psi(0,0)\psi^\dagger(x,t)\rangle_{T,+}.
\]

Now a few words about the organization of the paper.
In Section 2 we formulate the problem and summarize
the main results.
In Section 3 we obtain the determinant formulae for
the f\/ield form factors.
In Section 4 we obtain the Fredholm determinant representation
for the dynamical correlation functions.
In Section 5 we consider the completely integrable
dif\/ferential equation which describes
the 2-point dynamical
correlation functions with Neumann
boundary conditions.
In Section 6 we consider the special case that time $t=0$
and 
derive the Fredholm minor determinant representations
for the f\/inite-temperature f\/ields correlation functions.
We show that our Fredholm formulae coincides with the one which
has been obtained~\cite{kojima:K} at temperature $T=0$.

\setcounter{equation}{0}

\section{Formulation and Results}
The purpose of this section is to formulate the problem
and summarize the main results.
The Hamiltonian of our model is given by
\[
{\it H}=\int_0^L dx
\left(\partial_x\psi^\dagger\partial_x\psi
+c\psi^\dagger \psi^\dagger \psi \psi-h \psi^\dagger \psi\right)
+h_0\left(\psi^\dagger(0)\psi(0)-\psi^\dagger(L)\psi(L)\right).
\]
Here the f\/ields $\psi(x)$ and $\psi^\dagger(x)$ $(x \in {\bf R})$
are canonical Bose f\/ields given by
\[
[\psi(x),\psi^\dagger(y)]=\delta(x-y), \qquad
[\psi(x),\psi(y)]=[\psi^\dagger(x),\psi^\dagger(y)]=0,
\qquad (x,y \in {\bf R}),
\]
and $L>0$ is the size of box. 
The parameters $h>0$ and $h_0 \in {\bf R}$
represent the chemical potential and
the boundary chemical potential respectively.
We only consider the case of the coupling constant 
$c=\infty$, so-called ``impenetrable case''.
The Hamiltonian $\it H$ acts on the Fock space of the Bose f\/ields
def\/ined by the following relations between 
the Fock vacuum $\vert 0 \rangle$
and the Bose f\/ields:
\[
\langle 0 \vert \psi^{\dagger}(x)=0, \ \psi(x)\vert 0 \rangle=0,
\qquad
\langle 0 \vert 0 \rangle=1.
\]
A $N$-particle state vector $\vert \Psi_N\rangle$ is given by
\[
\vert \Psi_N\rangle
=\int_0^Ldz_1\ldots\int_0^Ldz_N\psi_N(z_1,\ldots,z_N)
\psi^\dagger(z_1)\ldots \psi^\dagger(z_N)\vert 0 \rangle,
\]
where the integrand $\psi_N(z_1,\ldots,z_N)$
is a $\bf C$-valued function.
The eigenvector problem:
${\it H}\vert \Psi_N \rangle
=E_N \vert \Psi_N \rangle$, $(E_N \in {\bf R})$,
is equivalent to the quantum mechanics problem def\/ined 
by the following
four conditions of the integrand function 
$\psi_N(z_1,\ldots,z_N)$.

\begin{enumerate}
\item
The wave function $\psi_N=\psi_N(z_1,\ldots,z_N)$ satisf\/ies
the free-particle Schr\"odinger equation in the case of variables
$0 < z_i \neq z_j < L$:
\[
\ba{l}
\ds -\sum_{j=1}^N \left(\frac{\partial}{\partial z_j}\right)^{2}
\psi_N(z_1,\ldots,z_N)=
E_N\cdot\psi_N(z_1,\ldots,z_N),
\vspace{3mm}\\
\ds (0< z_i\neq z_j < L, \ E_N \in {\bf R}).
\ea
\]

\item
The wave function $\psi_N$ is 
symmetric with respect to the variables:
\[
\psi_N(z_1,\ldots,z_N)=
\psi_N(z_{\sigma(1)},\ldots,z_{\sigma(N)}), \qquad (\sigma \in S_N).
\]

\item
The wave function $\psi_N$ satisf\/ies 
the integrable open boundary conditions:
\[
\ba{l}
\ds \left.\left(\frac{\partial}{\partial z_j}-h_0\right)
\psi_N\right|_{z_j=0}=0,
\vspace{3mm}\\
\ds \left.\left(\frac{\partial}{\partial z_j}+h_0\right)
\psi_N\right|_{z_j=L}=0, \qquad (j=1,\ldots,N).
\ea
\]
\item
The wave function $\psi_N$ vanishes whenever 
the coordinates coincide:
\[
\left.\psi_N(z_1,\ldots,z_i,\ldots,z_j,\ldots,z_N)\right|
_{z_i=z_j}=0.
\]
This condition corresponds to the condition: $c \to \infty$.
\end{enumerate}

The wave functions $\psi_N$ which satisfy 
the above four conditions were 
constructed~\cite{kojima:K}. They are parameterized by
the spectral parameters
\[
\ba{l}
\ds \psi_N(z_1,\ldots,z_N\vert\lambda_1,\ldots,\lambda_N)
\vspace{3mm} \\
\ds \qquad =Cons.\prod_{1\leq j < k \leq N}{\rm sgn}(z_j-z_k)
\det_{1 \leq j,k \leq N}
\left(\lambda_j \cos(\lambda_j z_k)+h_0 \sin(\lambda_j z_k)
\right).
\ea
\]
Here the function $\ds {\rm sgn}(x)=\frac{x}{\vert x \vert}$
and the spectral parameters $0\leq
\lambda_1<\lambda_2<\cdots<\lambda_N$ are determined by
the so-called Bethe Ansatz equations:
\be
\lambda_j=\frac{\pi}{L}I_j,
\qquad (I_j \in {\bf N}, \ j=1,2,\ldots,N).\label{kojima:BAE}
\ee
Because the coupling constant $c \to \infty$,
the Bethe Ansatz equations become simple.
The constant factor ``$Cons.$'' is determined by
\[
\langle \Psi_N(\lambda_1,\ldots,\lambda_N)\vert
\Psi_N(\lambda_1,\ldots,\lambda_N)\rangle = (2L)^N.
\]
The eigenvalue $E_N(\{\lambda\})$:
\[
{\it H}\ket = E_N(\{\lambda\})\ket,
\]
is given by
\[
E_N(\{\lambda\})=\sum_{j=1}^N(\lambda_j^2-h).
\]
We assume that the set
$\{ \vert \Psi_N(\lambda_1,\ldots,\lambda_N)\rangle \}
_{{\rm all}\{\lambda\}_N ~ N \in {\bf N}}$
is a basis of physical space of this model. Here the index
${\rm all}\{\lambda\}_N$
represents all the solutions of the Bethe Ansatz
equations~(\ref{kojima:BAE}).
This type assumption is usually called ``Bethe Ansatz''.
The following lemma is a foundation of our analysis.

\begin{lem} If the boundary condition $h_0$ takes the special
value $h_0=0, \infty$,
the eigenvectors $\vert \Psi(\{\lambda\})\rangle$
satisfy orthogonality relations
\be
\langle \Psi_N(\lambda_1,\ldots,\lambda_N)\vert
\Psi_N(\mu_1,\ldots,\mu_N)\rangle =(2L)^N
\prod_{j=1}^N \delta_{\lambda_j,\mu_j}, \qquad
(h_0=0,\infty).\label{kojima:orthogonal}
\ee
Here $\delta_{\lambda,\mu}$ is Kronecker Delta.
\end{lem}

To prove the above lemma, we have used 
the Bethe-Ansatz equations of 
the spectral parameters.
In the sequel we use the orthogonality relations of 
the eigenstates,
therefore we concentrate our attentions to 
the case of the special boundary
conditions: $h_0=0,\infty$.
The boundary conditions $h_0=0$ and $h_0=\infty$ are called
Neumann, Dirichlet, respectively.
In the sequel we use the following abberivations
\[
\vert \Psi_N(\lambda_1,\ldots,\lambda_N)\rangle_+
\quad {\rm for~Neumann}, \qquad
\vert \Psi_N(\lambda_1,\ldots,\lambda_N)\rangle_-
\quad {\rm for~Dirichlet}.
\]
The constant ``{\it Cons.}'' is given by
\[
Cons.=\left\{
\begin{array}{ll}
\ds \frac{2^N}{\sqrt{(1+\delta_{\lambda_1,0})N!}}
\left(\prod_{j=1}^N\lambda_j\right)^{-1},
&{\rm for~Neumann},
\vspace{3mm}\\
\ds \frac{1}{\sqrt{N!}}\left(\frac{2i}{h_0}\right)^N,&
{\rm for~Dirichlet}.
\end{array}\right.
\]

By using the orthogonal relations (\ref{kojima:orthogonal})
and the so-called ``Bethe Ansatz'',
we arrive at the completeness relation:
\be
id=\sum_{N=0}^{\infty} \; \sum_{{\rm all}\{\lambda\}_N}
\frac{\vert \Psi_N(\lambda_1,\ldots,\lambda_N)\rangle_{\epsilon}
~_{\epsilon}\langle\Psi_N(\lambda_1,\ldots,\lambda_N)\vert}
{_{\epsilon}\langle\Psi_N(\lambda_1,\ldots,\lambda_N)\vert
\Psi_N(\lambda_1,\ldots,\lambda_N)\rangle_{\epsilon}}.
\label{kojima:completeness}
\ee
The Bose f\/ields $\psi(x,t)$, $\psi^\dagger(x,t)$
are developed by the time $t$ by
\[
i\partial_t \psi =[\psi, H], \qquad
i\partial_t \psi^\dagger
=[\psi^\dagger, H].
\]
More explicitly the time dependence of 
the Bose f\/ields are written by
\[
\psi(x,t)=e^{iHt}\psi(x)e^{-iHt}, \qquad
\psi^\dagger(x,t)=e^{iHt}\psi^\dagger(x)e^{-iHt}.
\]
In this paper we are interested in 
the dynamical correlation functions
$\!\langle \psi(x_1,t_1)\psi^\dagger(x_2,t_2)\rangle_{\epsilon,T}\!\!$
def\/ined by the following way.
For the nonzero temperature $T>0$,
the dynamical correlation functions for $N$ state are def\/ined by
the summation of the every states:
\[
\ba{l}
\ds \langle \psi(x_1,t_1)\psi^\dagger(x_2,t_2)
\rangle_{\epsilon,N,T}
\vspace{3mm}\\
\ds \qquad =\left\{\sum_{{\rm all} \{\lambda\}_N}
\exp\left(-\frac{E_N(\{\lambda \})}{T}\right)
\right\}^{-1}
 \left\{\sum_{{\rm all} \{\lambda\}_N}
\exp\left(-\frac{E_N(\{\lambda \})}{T}\right)\right.
\vspace{3mm}\\
\ds \qquad \left. \times \frac{\bra \psi(x_1,t_1)
\psi^\dagger(x_2,t_2) \ket}
{\bra \Psi_N(\lambda_1,\ldots,\lambda_N)
\rangle_{\epsilon}}\right\},
\ea
\]
where the index $\epsilon=\pm$ represents 
the boundary conditions.
Now the index ``$+$''
represents the dynamical correlation functions 
with Neumann boundary
conditions.
The index ``$-$''
represents the dynamical correlation functions 
with Dirichlet boundary
conditions.
For the ground state case $T=0$,
the dynamical correlation functions for $N$ state are def\/ined by
the vacuum expectation value of the ground state:
\[
\ba{l}
\ds \langle \psi(x_1,t_1)\psi^\dagger(x_2,t_2)\rangle_{\epsilon,N,0}
\vspace{3mm}\\
\ds \qquad = \frac{\bra \psi(x_1,t_1)
\psi^\dagger(x_2,t_2) \ket}
{\bra \Psi_N(\lambda_1,\ldots,\lambda_N)\rangle_{\epsilon}},
\ea
\]
where the spectral parameters 
$(\lambda_1,\ldots,\lambda_N)$ are given by
\[
\lambda_j=
\left\{\begin{array}{ll}
\ds \frac{\pi}{L}(j-1), &{\rm for~Neumann},
\vspace{3mm}\\
\ds \frac{\pi}{L}j, &{\rm for~Dirichlet}.
\end{array}\right.
\]
In this paper we are interested in the thermodynamic limit
of the correlation functions, for the temperature $T\geq 0$
and the time $t_1,t_2 \in {\bf R}$.
For the nonzero temperature $T>0$, the dynamical 
correlation functions in the thermodynamic limit are def\/ined by
\[
\langle \psi(x_1,t_1)\psi^\dagger(x_2,t_2)\rangle_{\epsilon,T}
=\lim_{N,L \to \infty \atop{\frac{N}{L}=D(T)}}
\langle \psi(x_1,t_1)\psi^\dagger(x_2,t_2)\rangle_{\epsilon,N,T}.
\]
Here the density $\ds D(T)=\frac{N}{L}$ \cite{kojima:Y.Y.} is given by
\be
D(T)=\frac{1}{\pi}
\int_{0}^{\infty}\vartheta(\lambda)d\lambda,\label{kojima:def:Density}
\ee
where the Fermi weight $\vartheta(\lambda)$ is def\/ined
in~(\ref{kojima:def:Fermi}).
For the ground state $T=0$, the dynamical correlation function
in the thermodynamic limit is def\/ined by
\[
\langle \psi(x_1,t_1)\psi^\dagger(x_2,t_2)\rangle_{\epsilon,0}
=\lim_{N,L \to \infty \atop{\frac{N}{L}=D(0)}}
\langle \psi(x_1,t_1)\psi^\dagger(x_2,t_2)\rangle_{\epsilon,N,0}.
\]
Here the density $\ds D(0)=\frac{N}{L}$ can be chosen arbitrary.
In this paper we give the Fredholm determinant representations
for the dynamical correlation functions
$\langle \psi(x_1,t_1)\psi^\dagger(x_2,t_2)
\rangle_{\epsilon,T}$, $(T \geq 0 ,\epsilon =\pm)$.
Because the following relation hols:
\[
\langle \psi(x_1,0)\psi^\dagger(x_2,t_2-t_1)
\rangle_{\epsilon,T}
=\langle \psi(x_1,t_1)\psi^\dagger(x_2,t_2)
\rangle_{\epsilon,T},
\]
we only need one time parameter $t=t_2-t_1$,
to describe correlation functions.
In the sequel,
we use the abbreviation $t=t_2-t_1$.
Let us set 
\be
\tau(s|x,t)= its^2-ixs,\label{kojima:def:tau}
\ee
\be
G(x)=\frac{1}{2\pi}
\int_{-\infty}^{\infty}e^{\tau(s|x,t)}ds,
\label{kojima:def:G}
\ee
and
\be
\!\! P(\lambda \vert x_1,x_2)=e^{-\frac{1}{2}it \lambda^2}
\left\{e^{\tau(\lambda|x_1,t)}-\frac{2}{\pi}~{\rm P.V.}
\int_{-\infty}^{\infty}\frac{1}{s-\lambda}e^{\tau(s|x_1,t)}
\sin(x_2(s-\lambda))
ds\right\}.\label{kojima:kernel:P}
\ee
In Section 4 we derive the following formulae.

\begin{thm} In the thermodynamic limit
$N,L \to \infty$, such that $\ds \frac{N}{L}=D$,
the ground state dynamical correlation functions
have the Fredholm determinant representations
\[
\ba{l}
\ds \langle\psi(x_1,0)\psi^\dagger(x_2,t)\rangle_{\epsilon,0}
\vspace{3mm}\\
\ds \qquad =e^{-iht}
\left(G(x_1-x_2)+\epsilon G(x_1+x_2)+
\frac{1}{2\pi}\frac{\partial}{\partial \alpha}\right)
\left.\det\left(1-\frac{2}{\pi}
\widehat{V}_{\epsilon}-\alpha 
\widehat{A}_{\epsilon}\right)\right|_{\alpha=0},
\ea
\]
where the function $G(x)$ is given in (\ref{kojima:def:G}).
Here the integral operators $\widehat{V}_{\epsilon}$ and
$\widehat{A}_{\epsilon}$ are defined~by
\[
(\widehat{V}_{\epsilon}f)(\lambda)
=\int_{0}^q {V}_{\epsilon}(\lambda,\mu)f(\mu)d \mu,
\qquad
(\widehat{A}_{\epsilon}f)(\lambda)
=\int_{0}^q {A}_{\epsilon}(\lambda,\mu)f(\mu)d \mu,
\]
where the 
Fermi sphere $q=\pi D$ and the integral kernel are given by
Neumann or Dirichlet sum:
\[
V_{\epsilon}(\lambda,\mu)=
L(\lambda,\mu)
+\epsilon L(\lambda,-\mu),
\]
\[
A_{\epsilon}(\lambda,\mu)=\epsilon
\left(P(\lambda \vert x_1,x_2)+
\epsilon P(-\lambda \vert x_1,x_2)\right)
\left(P(\mu \vert x_2,x_1)+
\epsilon P(-\mu \vert x_2,x_1)\right).
\]
Here we have used the function $L(\lambda,\mu)$
defined in (\ref{kojima:kernel-L}) and
the function $P(\lambda \vert x_1,x_2)$
defined in (\ref{kojima:kernel:P}). Here we can choose the density
$D>0$ arbitrary.
\label{kojima:Th1}\end{thm}

We have succeeded to write the integral kernel 
by elementary functions:
\[
{\rm P.V.}
\int_{-\infty}^{\infty}\frac{e^{\tau(s|y,t)}}{s-\lambda}ds,
\]
and trigonometric functions.
In Section 4, we consider the f\/inite temperature case, too.

\begin{thm} In the thermodynamic limit:
$N,L \to \infty$, such that $\ds \frac{N}{L}=D(T)$~(\ref{kojima:def:Density}),
the finite temperature dynamical correlation functions
have the Fredholm determinant representations
\[
\ba{l}
\ds \langle\psi(x_1,0)\psi^\dagger(x_2,t)\rangle_{\epsilon,T}
 =e^{-iht}
\left(G(x_1-x_2)+\epsilon G(x_1+x_2)+
\frac{1}{2\pi}\frac{\partial}{\partial \alpha}\right)
\vspace{3mm}\\
\ds \qquad
\left. \times \det\left(1-\frac{2}{\pi}
\widehat{V}_{\epsilon,T}-\alpha 
\widehat{A}_{\epsilon,T}\right)\right|_{\alpha=0}.
\ea
\]
Here the temperature $T>0$ and
the integral operators $\widehat{V}_{\epsilon,T}$ and
$\widehat{A}_{\epsilon,T}$ are defined by
\[
(\widehat{V}_{\epsilon,T}f)(\lambda)
=\int_{0}^{\infty} {V}_{\epsilon}(\lambda,\mu)
\vartheta(\mu)f(\mu)d \mu,
\qquad
 (\widehat{A}_{\epsilon,T}f)(\lambda)
=\int_{0}^{\infty} {A}_{\epsilon}(\lambda,\mu)
\vartheta(\mu)f(\mu)d \mu,
\]
where the Fermi weight $\vartheta(\lambda)$ 
is given in (\ref{kojima:def:Fermi}).\label{kojima:Cor1}
\end{thm}

In Section 5, we derive the dif\/ferential equations
for correlation functions for the case $x_1=0$.
To describe $2n$ point dynamical correlation functions
for an impenetrabel Bose gas without boundaries,
N.~Slavnov~\cite{kojima:Sl} introduced a system of nonlinear
partial dif\/ferential equations,
which becomes the nonlinear Schr\"odinger equation
in the simplest case.
In this paper, we express the dynamical correlation functions
$\langle \psi(0,0)\psi^\dagger(x,t)\rangle_{T,+}$, $(T\geq 0)$
in terms of Slavnov's generalization of 
the nonlinear Schr\"odinger equations~\cite{kojima:Sl}.
For $x_1=0$ and Dirichlet boundary case:
\[
\langle \psi(0,0)\psi^\dagger(x_2,t)\rangle_{T,-}=0,
\]
because the wave functions become to zero.
We consider the case $x_1=0$
and Neumann boundary conditions.
First we consider $T=0$ case. Let us set
\be
\left(\widehat{W}f\right)(\lambda)
=\int_0^q W(\lambda,\mu)f(\mu)d\mu, \qquad (q=\pi D),
\label{kojima:def:W}
\ee
\be
W(\lambda,\mu)=\frac{\sin(x(\lambda-\mu))}{\lambda-\mu}+
\frac{\sin(x(\lambda+\mu))}{\lambda+\mu}.\label{kojima:kernel:W}
\ee

\begin{thm}
The correlation functions for an impenetrable Bose gas
with Neumann boundaries at the ground state 
are given by the following 
formulaes:
\[
\langle \psi(0,0)\psi^\dagger(x,t)\rangle_{0,+}
=2e^{-iht}\det\left(1-\frac{2}{\pi}\widehat{W}\right)
b_{1,4}\left(\begin{array}{cccc}
0&0&-x&x\\
0&0&t&t\end{array}\right).
\]
Here the integral operator $\widehat{W}$
is defined in (\ref{kojima:def:W}) and
the function $b_{1,4}$ is a component
of matrix $b$ defined in (\ref{kojima:def:b}).
\label{kojima:Th2}
\end{thm}

The matrix $b$ def\/ined in (\ref{kojima:def:b}) 
satisf\/ies a set of 
partial dif\/ferential equations introduced in~\cite{kojima:Sl}:
\be
\frac{\partial}{\partial t_j}L_k
-  \frac{\partial}{\partial y_k}M_j
+[L_k,M_j]=0, \qquad (1\leq j,k \leq 4).\label{kojima:Lax}
\ee
Here we have used
\[
L_j(\mu)=\mu P_j+[b,P_j], \qquad
M_j(\mu)=-\mu L_j(\mu)+\frac{\partial b}{\partial y_j},
\]
where we have used the matrix $P_j$ whose components
are def\/ined by
\be
(P_j)_{l,m}=i\delta_{l,j}\delta_{m,j}.\label{kojima:def:P}
\ee
The dif\/fential equations (\ref{kojima:Lax})
describe the logarithmic derivatives of the four point
correlation functions without boundaries, too:
\[
\langle \psi(y_1,t_1)\psi^\dagger(y_2,t_2)
\psi(y_3,t_3)\psi^\dagger(y_4,t_4)\rangle_{0}.
\]
In Section 5, we consider the f\/inite temperature case, too.
Let us consider the f\/inite temperature case $T>0$.
Let us set
\be
\left(\widehat{W}_Tf\right)(\lambda)
=\int_0^{\infty} W_T(\lambda,\mu)f(\mu)d\mu,
\label{kojima:def:WT}
\ee
\be
W_T(\lambda,\mu)=W(\lambda,\mu)\vartheta(\mu),
\label{kojima:kernel:WT}
\ee
where the kernel $W(\lambda,\mu)$ is def\/ined in (\ref{kojima:kernel:W})
and $\vartheta(\mu)$ is def\/ined in (\ref{kojima:def:Fermi}).

\begin{thm}
The correlation functions for an impenetrable Bose gas
with Neumann boundaries are given by the following 
formulaes:
\[
\langle \psi(0,0)\psi^\dagger(x,t)\rangle_{T,+}
=2e^{-iht}\det\left(1-\frac{2}{\pi}\widehat{W}_T\right)
b_{1,4}^T\left(\begin{array}{cccc}
0&0&-x&x\\
0&0&t&t\end{array}\right),
\]
Here the integral operator $\widehat{W}_T$
is defined in (\ref{kojima:def:WT}) and 
the function $b_{1,4}^T$ is a component of the matrix
$b^T$ defined in (\ref{kojima:def:bT}).
\label{kojima:Cor2}
\end{thm}

The matrix $b^T$ satisfy a set of dif\/fential equations
(\ref{kojima:Lax}), too. (We substitute $b$ to $b^T$.)

T. Kojima \cite{kojima:K}
derived the Fredholm minor determinants
formulae for the ground state correlation functions:
$\langle \psi(x_1)\psi^\dagger(x_2)\rangle_{0,\epsilon}$.
In Section 6 of this paper,
we consider the special case for the time $t=0$
of our Fredholm determinant formulae:
\[
\langle \psi(x_1,0)\psi^\dagger(x_2,0)\rangle_{T,\epsilon},
\qquad (\epsilon=\pm)
\]
and derive the Fredholm minor determinant formulae for temperature
$T\geq 0$.
This  Fredholm minor determinant formulae for $T=0$
coincide with the one which
has been obtained~\cite{kojima:K}. Let us set
\be
\left(\hat{\theta}_{\epsilon,T}^{(y_1,y_2)}f\right)(\xi)
=\int_{0}^{\infty}
\left(({\rm E}(y_1-\xi')+{\rm E}(y_2-\xi'))
\theta_{\epsilon,T}(\xi,\xi')\right)f(\xi')d\xi',
\label{kojima:kertheta}
\ee
where
\[
\theta_{\epsilon,T}(\xi,\eta)
=\int_0^{\infty}\vartheta(\nu)
\left\{\cos((\xi-\eta)\nu)+\epsilon \cos((\xi+\eta)\nu)
\right\}d\nu,
\]
and $\vartheta(\lambda)$ is def\/ined in (\ref{kojima:def:Fermi}).
Here ${\rm E}(\xi)$ represents the step function
\[
{\rm E}(\xi)=\left\{
\begin{array}{ll}1, & {\rm for} \quad  \xi \geq 0, \\
0, & {\rm for} \quad  \xi < 0.
\end{array}\right.
\]

\begin{thm} For the temperature $T\geq 0$,
the field correlation functions 
have the first Fredholm minor determinants representations
\[
\langle \psi(x_1)\psi^{\dagger}(x_2)\rangle_{\epsilon,T}
=\frac{1}{2}
\det\left(1-\frac{2}{\pi}\widehat{\theta}
^{(x_1,x_2)}_{\epsilon,T}
\left|\begin{array}{c}
x_2 \\ x_1
\end{array}\right.
\right),
\]
where the integral operator 
$\widehat{\theta}^{(x_1,x_2)}_{\epsilon,T}$ is 
defined in (\ref{kojima:kertheta}).
\label{kojima:Th3}
\end{thm}

\begin{thm}\cite{kojima:K}
The ground state correlation functions have the
first Fredholm minor determinant formulae
\[
\langle \psi(x_1)\psi^{\dagger}(x_2)\rangle_{\epsilon,0}
=\frac{1}{2}
\det\left(1-\frac{2}{\pi}\widehat{K}^{(x_1,x_2)}_{\epsilon}
\left|\begin{array}{c}
x_2 \\ x_1
\end{array}\right.
\right).
\]
Here the integral operator is defined by
\[
\left(\hat{K}_{\epsilon}^{(x_1,x_2)}f\right)(\xi)
=\int_{x_1}^{x_2}K_{\epsilon}(\xi,\xi')f(\xi')d\xi',
\]
where
\[
K_{\epsilon}(\xi,\eta)=\frac{\sin D(\xi-\eta)}{\xi-\eta}
+\epsilon
\frac{\sin D(\xi+\eta)}{\xi+\eta}.
\]
Here the density $\ds D=\frac{N}{L}$ can be chosen arbitrary.
\label{kojima:Cor3}
\end{thm}

\setcounter{equation}{0}

\section{Form Factors}

{\advance\topsep-2pt
The purpose of
this section is to derive the determinant formulae 
for the form factors.
First we prepare a lemma.
\begin{lem} For the sequences
$\{f_{j,k}\}_{j=1,\ldots,N+1,~k=1,\ldots,N}$ and
$\{g_{j}\}_{j=1,\ldots,N+1}$,
the following holds:
\be \label{kojima:det:form_eq}
\!\! \left.\sum_{\sigma \in S_{N+1}}{\rm sgn}\;\sigma
f_{\sigma(N+1)}\prod_{j=1}^N g_{\sigma(j),j}=
\left(f_{N+1}+\frac{\partial}{\partial \alpha}\right)
\det_{1 \leq j,k \leq N}
\left(g_{j,k}-\alpha f_j \cdot g_{N+1,k}\right)
\right|_{\alpha=0} \!\!\!.
\ee
\label{kojima:det:form}
\end{lem}

\noindent
{\sl Proof.} Consider the coset decomposition:
\[
S_{N+1}=S_N(N+1) \cup S_N(N)\cdot(N,N+1) 
\cup \cdots \cup
 S_N(1)\cdot(1,N+1),
\]
where $S_N(j)$ is permutations of $(1,\ldots,j-1,N+1,j+1,
\ldots,N)$.
Rewrite the left side of the equation (\ref{kojima:det:form_eq}) 
with respect to the coset decomposition:
\[
\ba{l}
\ds (L.H.S.)=\sum_{j=1}^N g_j \sum_{\tau \in S_N(j)}{\rm sgn}(\tau
\cdot (j,N+1))\prod_{k=1 \atop{k \neq j}}^N f_{\tau(k),k}\cdot
f_{\tau(N+1),j}
\vspace{3mm} \\
\ds \qquad
+g_{N+1}\sum_{\tau \in S_N(N+1)}{\rm sgn}(\tau \cdot(N+1,N+1))
\prod_{k=1}^N f_{\tau(k),k}=(R.H.S.)
\ea
\]

\subsection*{Q.E.D.}
Now let us consider the f\/ield form factor:
\[
\ba{l}
_{\epsilon}\langle \Psi_{N+1}
(\lambda_1,\ldots,\lambda_{N+1})\vert
\psi^\dagger(x)\vert \Psi_N(\mu_1,\ldots,\mu_N)\rangle_{\epsilon}
\vspace{3mm}\\
\ds \qquad =\sqrt{N+1}\int_0^Ldz_1 \ldots \int_0^Ldz_N
\psi_{N+1}^*(z_1,\ldots,z_N,x\vert \lambda_1,\ldots,\lambda_{N+1})
\vspace{3mm}\\
\ds \qquad \times\psi_{N}
(z_1,\ldots,z_N\vert \mu_1,\ldots,\mu_{N})
\vspace{3mm}\\
\ds \qquad =\frac{1}{\sqrt{(1+\delta_{\lambda_1,0})(1+\delta_{\mu_1,0})}}
\sum_{\sigma \in S_{N+1}}{\rm sgn}\; \sigma
(e^{-i\lambda_{\sigma(N+1)}x}+\epsilon
e^{i\lambda_{\sigma(N+1)}x})
\vspace{3mm}\\
\ds \qquad \times
\prod_{j=1}^N\left\{
\int_0^L dz ~{\rm sgn}\; (z-x)
(e^{-i\lambda_{\sigma(j)}z}+\epsilon e^{i\lambda_{\sigma(j)}z})
(e^{i\mu_{j}z}+\epsilon e^{-i\mu_{j}z})
\right\}.
\ea
\]
To derive the third line, we have used a simple fact:
\[
\sum_{\sigma,\tau \in S_N}{\rm sgn}\;\sigma \tau
\prod_{j=1}^N f_{\sigma(j),\tau(j)}
=N! \sum_{\sigma \in S_N}{\rm sgn}\;\sigma \prod_{j=1}^N f_{\sigma(j),j}.
\]
Using lemma \ref{kojima:det:form},
we arrive at the determinant formulae for the form factors.

\begin{lem} 
The field form factors have the determinant formula
\[
\ba{l}
\ds _{\epsilon}\langle \Psi_{N+1}
(\lambda_1,\cdots,\lambda_{N+1})\vert
\psi^\dagger(x)\vert \Psi_N(\mu_1,\ldots,\mu_N)\rangle_{\epsilon}
=\left(C_{\epsilon}(x \vert \lambda_{N+1})+
\frac{\partial}{\partial \alpha}\right)
\vspace{3mm}\\
\ds \left. \qquad \times \det_{1 \leq j,k \leq N}
\left(I_{\epsilon}(x \vert \lambda_j,\mu_k)-
\alpha C_{\epsilon}(x \vert \lambda_j)
I_{\epsilon}(x \vert \lambda_{N+1},\mu_k)\right)\right|_{\alpha=0}.
\ea
\]
Here we have used 
\be
C_{\epsilon}(x \vert \lambda)=
\frac{1}{\sqrt{1+\delta_{\lambda,0}}}(e^{-i\lambda x}+
\epsilon e^{i\lambda x}),
\label{kojima:def:C}
\ee
\be\label{kojima:def:I}
\ba{l}
\ds I_{\epsilon}(x \vert \lambda,\mu)=
\frac{1}{\sqrt{(1+\delta_{\lambda,0})(1+\delta_{\mu,0})}}\left\{
\frac{4}{\lambda-\mu}\sin(x(\lambda-\mu))\right.
\vspace{3mm}\\
\ds \qquad +\left.\epsilon\frac{4}{\lambda+\mu}\sin(x(\lambda+\mu))
-2L(\delta_{\lambda,\mu}+\epsilon \delta_{\lambda,0}\delta_{\mu,0})
\right\}.
\ea
\ee
\label{kojima:lem:form}
\end{lem}
We can write the f\/ield form factors without using integrals.

From the relation $\psi^\dagger(x,t)=e^{iHt}\psi^\dagger(x)e^{-iHt}$,
the dynamical form factors are given by
\be \label{kojima:timform}
\ba{l}
_{\epsilon}\langle \Psi_{N+1}
(\lambda_1,\ldots,\lambda_{N+1})\vert
\psi^\dagger(x,t)\vert \Psi_N(\mu_1,\ldots,\mu_N)
\rangle_{\epsilon}
\vspace{3mm}\\
\ds \qquad =\exp\left\{it\left(-h+\sum_{j=1}^{N+1}\lambda_j^2
-\sum_{j=1}^N\mu_j^2\right)\right\}
\vspace{3mm}\\
\ds \qquad \times  \;
_{\epsilon}\langle \Psi_{N+1}
(\lambda_1,\ldots,\lambda_{N+1})\vert
\psi^\dagger(x)\vert \Psi_N(\mu_1,\ldots,\mu_N)\rangle_{\epsilon}.
\ea
\ee

\setcounter{equation}{0}

\section{Correlation Functions}
The purpose of this section is to derive 
the Fredholm determinant formulas
of the dynamical correlation functions
$\langle\psi(x_1,t_1)\psi^\dagger(x_2,t_2)\rangle_{\epsilon,T}$.
First we consider the vacuum expectation values of 
f\/ields operators.
Using the completeness relation (\ref{kojima:completeness}),
the vacuum expectation values of two f\/ields are given by
\[
\ba{l}
\ds \frac{_\epsilon\langle \Psi_N(\mu_1,\ldots,\mu_N)\vert
\psi(x_1,t_1)\psi^\dagger(x_2,t_2)
\vert \Psi_N(\mu_1,\ldots,\mu_N)\rangle_\epsilon}
{_\epsilon\langle \Psi_N(\mu_1,\ldots,\mu_N)\vert
\Psi_N(\mu_1,\ldots,\mu_N)\rangle_\epsilon}
\vspace{3mm}\\
\ds \quad =
\sum_{{\rm all}\{\lambda\}_{N+1}}
\frac{_\epsilon\langle \Psi_N(\{\mu\})\vert
\psi(x_1,t_1)\vert \Psi_{N+1}(\{\lambda\})\rangle_{\epsilon}
~_{\epsilon}\langle \Psi_{N+1}(\{\lambda\}) \vert 
\psi^\dagger(x_2,t_2)\vert \Psi_N(\{\mu\})\rangle_\epsilon}
{_{\epsilon}\langle \Psi_N(\{\mu\})\vert
\Psi_N(\{\mu\})\rangle_{\epsilon}
~_{\epsilon}\langle \Psi_{N+1}(\{\lambda\})\vert
\Psi_{N+1}(\{\lambda\})\rangle_{\epsilon}}.
\ea\!
\]
Using the equation (\ref{kojima:timform}) and the following relations:
\[
_{\epsilon}\langle \Psi_N(\{\mu\})\vert\psi(x_1,t_1)
\vert \Psi_{N+1}(\{\lambda\})\rangle_{\epsilon}
=_{\epsilon}\langle \Psi_{N+1}(\{\lambda\})
\vert\psi^\dagger(x_1,t_1)
\vert \Psi_{N}(\{\mu\})\rangle_{\epsilon}^{*},
\]
\[
_{\epsilon}\langle \Psi_N(\{\lambda\})\vert
\Psi_N(\{\lambda\})\rangle_{\epsilon}
=(2L)^N,
\]
we obtain
\[
\ba{l}
\ds \frac{1}{(N+1)!}
\left(\frac{1}{2L}\right)^{2N+1}
e^{-i(t_2-t_1)(h+\sum_{j=1}^N\mu_j^2)}
\sum_{\lambda_1 \in \frac{\pi}{L}{\bf N}}\cdots
\sum_{\lambda_{N+1} \in \frac{\pi}{L}{\bf N}}
e^{i(t_2-t_1)\sum_{j=1}^{N+1}\lambda_j^2}
\vspace{3mm}\\
\ds \qquad \times \; _{\epsilon}\langle
\Psi_{N+1}(\{\lambda\})\vert\psi^\dagger(x_1) 
\vert \Psi_{N}(\{\mu\})\rangle_{\epsilon}^{*}
~_{\epsilon}\langle \Psi_{N+1}(\{\lambda\})\vert\psi^\dagger(x_2)
\vert \Psi_{N}(\{\mu\})\rangle_{\epsilon}.
\ea
\]
The translation invariance of time holds
\[
\ba{l}
\ds _\epsilon\langle \Psi_N(\{\mu\})\vert
\psi(x_1,t_1)\psi^\dagger(x_2,t_2)
\vert \Psi_N(\{\mu\})\rangle_\epsilon
\vspace{3mm}\\
\ds \qquad =\;
_\epsilon\langle \Psi_N(\{\mu\})\vert
\psi(x_1,0)\psi^\dagger(x_2,t_2-t_1)
\vert \Psi_N(\{\mu\})\rangle_\epsilon.
\ea
\]
In the sequel we set the abberiviation $t=t_2-t_1$.
Remember a following simple fact.
{\it For sequences} $\{f_{j_1,\ldots,j_n}\}
_{j_1,\ldots,j_n \in I},
\{g_{j_1,\ldots,j_n}\}
_{j_1,\ldots,j_n \in I}$,
({\it I}: some index set),
{\it the following holds}
\[
\sum_{j_1,\ldots,j_n \in I}
({\rm Sym}~f)_{j_1,\ldots,j_n}
({\rm Sym}~g)_{j_1,\ldots,j_n}
=\sum_{j_1,\ldots,j_n \in I}
f_{j_1,\ldots,j_n}~
({\rm Sym}~g)_{j_1,\ldots,j_n}.
\]
{\it Here we have used}
\[
({\rm Sym}~f)_{j_1,\ldots,j_n}=\frac{1}{n!}
\sum_{\sigma \in S_n}f_{j_{\sigma(1)},\ldots,j_{\sigma(n)}}.
\]

The form factors have the determinant formulae in lemma 
\ref{kojima:lem:form} and
\[
_{\epsilon}\langle \Psi_{N+1}(\{\lambda\})\vert\psi^\dagger(x)
\vert \Psi_{N}(\{\mu\})\rangle_{\epsilon}
=\sum_{\sigma \in S_{N+1}}
C_{\epsilon}(x \vert \lambda_{\sigma(N+1)})
\prod_{j=1}^N I_{\epsilon}(x \vert \lambda_{\sigma(j)},\mu_j),
\]
We obtain
\[
\ba{l}
\ds e^{-it\left(h+\sum\limits_{j=1}^N\mu_j^2
\right)}\left(\frac{1}{2L}\right)^{2N+1}
\vspace{3mm}\\
\ds \qquad \times \sum_{\lambda_1 \in \frac{\pi}{L}{\bf N}}
\cdots
\sum_{\lambda_{N+1} \in \frac{\pi}{L}{\bf N}}
\left(e^{it\lambda_{N+1}^2}C_{\epsilon}^*(x_1 \vert \lambda_{N+1})
C_{\epsilon}(x_2 \vert \lambda_{N+1})
+\frac{\partial}{\partial \alpha}\right)
\vspace{3mm}\\
\ds \qquad \times \left.\det_{1 \leq j,k \leq N}\left(
e^{it\lambda_j^2}I_{\epsilon}(x_1 \vert \lambda_j, \mu_k)
I_{\epsilon}(x_2 \vert \lambda_j, \mu_j)-\alpha
e^{it\lambda_j^2}C^*_{\epsilon}(x_1 \vert \lambda_j)
I_{\epsilon}(x_2 \vert \lambda_j, \mu_j)\right. \right.
\vspace{3mm}\\
\ds \qquad \times \left. \left.
e^{it\lambda_{N+1}^2}C_{\epsilon}(x_2 \vert \lambda_{N+1})
I_{\epsilon}(x_1 \vert \lambda_{N+1}, \mu_k)
\right)\right|_{\alpha=0}.
\ea
\]
The $j$ th line of the above matrix only depends on $\lambda_j$
not on $\lambda_k,(k \neq j)$, therefore 
we can insert the summations
$\sum_{\lambda_1}\cdots \sum_{\lambda_{N+1}}$
 into the matrix. Now we arrive at the following.

\begin{prop} The vacuum expectation values of two fields have
the determinant formulas
\[
\ba{l}
\ds \frac{_\epsilon\langle \Psi_N(\mu_1,\ldots,\mu_N)\vert
\psi(x_1,0)\psi^\dagger(x_2,t)
\vert \Psi_N(\mu_1,\ldots,\mu_N)\rangle_\epsilon}
{_\epsilon\langle \Psi_N(\mu_1,\ldots,\mu_N)\vert
\Psi_N(\mu_1,\ldots,\mu_N)\rangle_\epsilon}
\vspace{3mm}\\
\ds \qquad = e^{-ith} 
\left( \frac{1}{2L}\sum_{s \in \frac{\pi}{L}{\bf N}}
\epsilon e^{it s^2}C_{\epsilon}(x_1 \vert s)
C_{\epsilon}(x_2 \vert s)
+\frac{\partial}{\partial \alpha}\right)
\vspace{3mm}\\
\ds \qquad \times \det_{1 \leq j,k \leq N}\left(
\left(\frac{1}{2L}\right)^2
\sum_{s \in \frac{\pi}{L}{\bf N}}
e^{it s^2}J_{\epsilon}(x_1 \vert s, \mu_k)
J_{\epsilon}(x_2 \vert s, \mu_j)\right.
\ea
\]
\[
\ba{l}
\ds \qquad - \alpha \epsilon \frac{1}{2L}
\left(\frac{1}{2L}
\sum_{s \in \frac{\pi}{L}{\bf N}}
e^{it s^2}C_{\epsilon}(x_1 \vert s)
J_{\epsilon}(x_2 \vert s, \mu_j)\right)
\vspace{3mm}\\
\ds \qquad \left. \left. \times 
\left(\frac{1}{2L}
\sum_{s \in \frac{\pi}{L}{\bf N}}
e^{it s^2}C_{\epsilon}(x_2 \vert s)
J_{\epsilon}(x_1 \vert s, \mu_k)\right)
\right)\right|_{\alpha=0}.
\ea
\]
Here we have used 
\[
J_{\epsilon}(x\vert s ,\mu)=e^{-\frac{1}{2}it\mu^2}
I_{\epsilon}(x\vert s,\mu).
\]
and functions $C_{\epsilon}(x \vert s)$
 and $I_{\epsilon}(x\vert s, \mu)$
are defined in (\ref{kojima:def:C}) and (\ref{kojima:def:I}), respectively.
\end{prop}

The size of the above matrix depends on the state number $N$,
however, the element of the matrix does not depend on $N$.
By calculations, we obtain
\be \label{kojima:tdl1}
\ba{l}
\ds \frac{1}{2L}\sum_{s \in \frac{\pi}{L}{\bf N}}
\epsilon e^{it s^2}C_{\epsilon}(x_1 \vert s)
C_{\epsilon}(x_2 \vert s)
\vspace{3mm}\\
\ds \qquad =\frac{1}{2L}\sum_{s \in \frac{\pi}{L}{\bf Z}}
e^{it s^2-is(x_1-x_2)}
+\epsilon
\frac{1}{2L}\sum_{s \in \frac{\pi}{L}{\bf Z}}
e^{it s^2-is(x_1+x_2)},
\ea
\ee
\be \label{kojima:tdl2}
\ba{l}
\ds \frac{1}{2L}
\sum_{s \in \frac{\pi}{L}{\bf N}}e^{it s^2}
C_{\epsilon}(x_1 \vert s)
J_{\epsilon}(x_2 \vert s, \mu)
 =\frac{e^{-\frac{1}{2}it\mu^2}}{\sqrt{1+\delta_{\mu,0}}}
\Biggl[
\Biggl\{
\frac{2}{\pi}\left(\frac{\pi}{L}\right)
\vspace{3mm}\\
\ds \qquad \times \sum_{s \in \frac{\pi}{L}{\bf Z}}
\frac{e^{it s^2-is x_1}}
{s-\mu}\sin(x_2(s-\mu))
-e^{it\mu^2-i\mu x_1}\Biggr\}+\epsilon
\{\mu \leftrightarrow (-\mu)\}\Biggr],
\ea
\ee
and
\be \label{kojima:tdl3}
\ba{l}
\ds \left(\frac{1}{2L}\right)^2
\sum_{s \in \frac{\pi}{L}{\bf N}}
e^{it s^2}J_{\epsilon}(x_1 \vert s, \mu)
J_{\epsilon}(x_2 \vert s, \lambda) 
=\delta_{\lambda,\mu}-\frac{2}{\pi}\left(\frac{\pi}{L}\right)
\frac{e^{-\frac{1}{2}it(\lambda^2+\mu^2)}}
{\sqrt{(1+\delta_{\lambda,0})(1+\delta_{\mu,0})}}
\vspace{3mm}\\
\ds \qquad \times \Biggl[\frac{1}{\lambda-\mu}
\Biggl\{
e^{it\lambda^2}\sin(x_1(\lambda-\mu))+
e^{it\mu^2}\sin(x_2(\lambda-\mu))
\vspace{3mm}\\
\ds \qquad - \frac{2}{\pi}\left(\frac{\pi}{L}\right)
\sum_{s \in \frac{\pi}{L}{\bf Z}}
e^{it s^2}\left(\frac{1}{s-\lambda}-\frac{1}{s-\mu}\right)
\sin((s-\mu)x_1)\sin((s-\lambda)x_2)\Biggr\}
\vspace{3mm}\\
\ds \qquad +\epsilon \frac{1}{\lambda+\mu}\{\mu 
\leftrightarrow (-\mu)\}\Biggr].
\ea
\ee
It is straightforwards to take the thermodynamic limit
of the right hand side of the equations (\ref{kojima:tdl1}),
(\ref{kojima:tdl2}) and (\ref{kojima:tdl3}).
We arrive at Theorem \ref{kojima:Th1}.
Next we consider the f\/inite temperature thermodynamics.
By statistical mechanics arguments,
at temperature $T>0$,
the thermodynamic equilibrium
distribution of the spectral parameters
is given by the Fermi weight $\vartheta(\lambda)$ 
(\ref{kojima:def:Fermi}):
\[
\lim \left(\frac{\pi}{L}\right)
\frac{1}{\lambda_{j+1}-\lambda_j}=
\vartheta(\lambda_j).
\]
Therefore the density is given by
\[
D(T)=\frac{N}{L}=\frac{1}{\pi}
\int^{\infty}_0 \vartheta(\lambda)d\lambda.
\]
Now we arrive at Theorem \ref{kojima:Cor1}.

} 

\setcounter{equation}{0}

\section{Dif\/ferential equations}

In this section we will study the most interesting case
$\langle \psi(0,0)\psi^\dagger(x,t)\rangle_{\epsilon,T}$,
which only appears in open boundary model.
For Dirichlet boundary case $\epsilon=-$,
\mbox{$\langle \psi(0,0)\psi^\dagger(x,t)\rangle_{-,T}=0,\hspace{-2.5pt}$}
because the wave functions become zero.
We will consider Neumann boundary case
$\langle \psi(0,0)\psi^\dagger(x,t)\rangle_{+,T}$ 
and derive the dif\/ferentaial equations which
describe the dynamical correlation functions.

\subsection{Preparations}

First we will consider the zero temperature 
and general $x_1,x_2$ case.

By using the relation:
\[
\left.\frac{\partial}{\partial \alpha}
\det\left(1-\frac{2}{\pi}\hat{V}_{\epsilon}-\alpha
\widehat{A}_{\epsilon}
\right)\right|_{\alpha=0}
=-\det\left(1-\frac{2}{\pi}\hat{V}_{\epsilon}\right)
{\rm Tr}\left(\left(1-\frac{2}{\pi}\hat{V}_{\epsilon}\right)
^{-1}\widehat{A}_{\epsilon}\right)
\]
we obtain the following formulae
\[
\ba{l}
\ds \langle\psi(x_1,0)\psi^\dagger(x_2,t)\rangle_{\epsilon,0}
=e^{-iht}
\det\left(1-\frac{2}{\pi}\hat{V}_{\epsilon}\right)
\vspace{3mm}\\
\ds \qquad \times \left(G(x_1-x_2)+\epsilon G(x_1+x_2)
-\frac{1}{2\pi}
{\rm Tr}\left(\left(1-\frac{2}{\pi}\hat{V}_{\epsilon}\right)
^{-1}\widehat{A}_{\epsilon}\right)\right).
\ea
\]
Def\/ine the integral operator $\widehat{R}_{\epsilon}$ by
\[
\left(\widehat{R}_{\epsilon}f\right)(\lambda)
=\int_0^q R_{\epsilon}(\lambda,\mu) f(\mu) d\mu.
\]
The kernel function 
$R_{\epsilon}(\lambda,\mu)$ is characterized by
the following integral equation:
\[
\left(1-\frac{2}{\pi}\widehat{V}_{\epsilon}\right)
\left(1+\frac{2}{\pi}\widehat{R}_{\epsilon}\right)=1.
\]
Def\/ine the integral operator $\widehat{S}$
and $\widehat{L}$ by
\[
\left(\widehat{S}f\right)(\lambda)
=\int_{-q}^q S(\lambda,\mu) f(\mu) d\mu, \qquad 
\left(\widehat{L}f\right)(\lambda)
=\int_{-q}^q L(\lambda,\mu) f(\mu) d\mu,
\]
where kernel function $L(\lambda,\mu)$ is def\/ined in 
(\ref{kojima:kernel-L}).
The kernel function 
$S(\lambda,\mu)$ is characterized by
the following integral equation:
\[
\left(1-\frac{2}{\pi}\widehat{L}\right)
\left(1+\frac{2}{\pi}\widehat{S}\right)=1.
\]

\begin{lem} The kernel functions
are related by the following linear relation
\[
R_{\epsilon}(\lambda,\mu)=S(\lambda,\mu)+\epsilon S(\lambda,-\mu).
\] 
\label{kojima:lem:kernel}
\end{lem}

\noindent
{\sl Proof.} The following characteristic relation holds:
\[
S(\lambda,\mu)-\frac{2}{\pi}\int_0^q 
\left(L(\lambda,\nu)S(\nu,\mu)+
L(\lambda,-\nu)S(-\nu,\mu)\right)
d\nu=L(\lambda,\mu).
\]
Using the relations $\epsilon^2=1$, $(\epsilon=\pm)$
and $L(\lambda,-\mu)=L(-\lambda,\mu)$,
we obtain the following characteristic relation:
\[
\ba{l}
\ds \left(S(\lambda,\mu)+\epsilon S(\lambda,-\mu)\right)
-\frac{2}{\pi}\int_0^q
\left(S(\lambda,\nu)+\epsilon S(\lambda,-\nu)\right)
\left(L(\nu,\mu)+\epsilon L(\nu,-\mu)\right)d\nu
\vspace{3mm}\\
\ds \qquad 
=L(\lambda,\mu)+\epsilon L(\lambda,-\mu).
\ea
\]

\subsection*{Q.E.D.}

By using lemma \ref{kojima:lem:kernel}, we obtain
\[
\ba{l}
\ds {\rm Tr}\left(\left(1-\frac{2}{\pi}\hat{V}_{\epsilon}\right)
^{-1}\widehat{A}_{\epsilon}\right)=
{\rm Tr}\left(\left(1+\frac{2}{\pi}\widehat{R}_{\epsilon}\right)
\widehat{A}_{\epsilon}\right)
\vspace{3mm}\\
\ds \qquad ={\rm Tr}\left(\left(1+\frac{2}{\pi}\widehat{S}\right)
\widehat{U}\right)
+\epsilon~{\rm Tr}\left(\left(1+\frac{2}{\pi}\widehat{S}\right)
\widehat{U}\widehat{Asy}\right).
\ea
\]
Here we have used
\be
\left(\widehat{U}f\right)(\lambda)
=\int_{-q}^q U(\lambda,\mu)f(\mu)d\mu,
\qquad {\rm where} \quad 
U(\lambda,\mu)=
P(\lambda|x_1,x_2)P(\mu|x_2,x_1).
\label{kojima:def:U}
\ee
Here we have used
\[
\left(\widehat{Asy}f\right)(\lambda)
=f(-\lambda).
\]
In the sequel of this section
we will consider the special case that
$x_1=0$, $x_2=x$ and $\epsilon=+$.
The following simplication occurs:
\[
\left.L(\lambda,\mu)\right|_{x_1=0,x_2=x}
=e^{\frac{1}{2}it(-\lambda^2+\mu^2)}
\frac{\sin(x(\lambda-\mu))}{\lambda-\mu}.
\]
Therefore
$\ds \left.\det\left(1-\frac{2}{\pi}\widehat{V}_{+}
\right)\right|_{x_1=0,x_2=x}$
dose not depend on time variable $t$:
\[
\left.\det\left(1-\frac{2}{\pi}\widehat{V}_{+}\right)
\right|_{x_1=0,x_2=x}=
\det\left(1-\frac{2}{\pi}\widehat{W}\right),
\]
where the operator $\widehat{W}$ is def\/ined in (\ref{kojima:def:W}).
By using the relation:
\[
P(-\lambda|x_1,x_2)=P(\lambda|-x_1,x_2),
\]
we obtain the simplication:
\[
{\rm Tr}\left(\left(1-\frac{2}{\pi}\hat{V}_{+}\right)
^{-1}\widehat{A}_{+}\right)=
2\left.{\rm Tr}\left(\left(1+\frac{2}{\pi}\widehat{S}\right)
\widehat{U}\right)\right|_{x_1=0,x_2=x}.
\]
We arrive at formulae
\[
\ba{l}
\ds \langle\psi(0,0)\psi^\dagger(x,t)\rangle_{+,0}
\vspace{3mm}\\
\ds \qquad =2 e^{-iht}
\det\left(1-\frac{2}{\pi}\hat{W}\right)
\left.\left(G(x)
-\frac{1}{2 \pi}
{\rm Tr}\left(\left(1+\frac{2}{\pi}\widehat{S}\right)
\widehat{U}\right)\right)\right|
_{x_1=0,x_2=x}.
\ea
\]

\subsection{Dif\/ferential Equations}
In this section we will f\/ind the partial
dif\/ferential equations of variables $t$ and $x$.
By discussion in the previous subsection,
it is enough to consider the factor:
\[
G(x_1+x_2)-\frac{1}{2\pi}
{\rm Tr}\left(\left(1+\frac{2}{\pi}\widehat{S}\right)
\widehat{U}\right).
\]
It is convenient to consider the problem
in more general situation.
We introduce the auxiliary functions $G_p(\lambda)$ and
the auxiliary vectors
$e_p^L(\lambda)$ and $e_p^R(\mu)$ def\/ined by
\[
G_p(\lambda)=\frac{1}{2\pi}~{\rm P.V.}
\int_{-\infty}^{\infty}\frac{1}{s-\lambda}
e^{\tau(s|y_{2p}-y_{2p-1}~,t_{2p}-t_{2p-1})}ds,
\]
\[
e_p^L(\lambda)=\left(
\begin{array}{cc}
-e^{it_{2p-1}\lambda^2-iy_{2p-1}\lambda}&
e^{it_{2p-1}\lambda^2-iy_{2p-1}\lambda}G_p(\lambda)
\end{array}
\right),
\]
\[
e_p^R(\mu)=\frac{2}{\pi}\left(
\begin{array}{c}
e^{-it_{2p}\mu^2+iy_{2p}\mu}G_p(\mu)
\vspace{1mm}\\
e^{-it_{2p}\mu^2+iy_{2p}\mu}
\end{array}
\right).
\]
Let us set the integral operators
$\widehat{K}_p$ by
\[
\left(\widehat{K}_pf\right)(\lambda)=
\int_{-\infty}^\infty K_p(\lambda,\mu)f(\mu)d\mu,
\]
where we have used the kernel def\/ined by
\[
K_p(\lambda,\mu)=\frac{\pi}{2}\frac{1}{\lambda-\mu}
e_p^L(\lambda)e_p^R(\mu).
\]
Let us set
\[
E^L(\lambda)=
\left(E^L_1(\lambda)~E^L_2(\lambda)~
E^L_3(\lambda)~E^L_4(\lambda)
\right)=\left(
\begin{array}{cc}
e_1^L(\lambda)&
\left(\left(1+\frac{2}{\pi}\widehat{K}_1\right)e_2^L\right)
(\lambda)
\end{array}
\right),
\]
\[
E^R(\mu)=
\left(
\begin{array}{c}
E^R_1(\mu)
\vspace{1mm}\\
E^R_2(\mu)
\vspace{1mm}\\
E^R_3(\mu)
\vspace{1mm}\\
E^R_4(\mu)
\end{array}
\right)=
\left(\begin{array}{c}
\left(e_1^R \left(1+\frac{2}{\pi}\widehat{K}_2\right)\right)
(\mu)
\vspace{1mm}\\
e_2^R(\mu)
\end{array}
\right).
\]
By direct calculations we obtain 
the closed dif\/ferential equations
of the above vectors:
\[
\frac{\partial}{\partial y_j}E^R(\mu)
=l_j(\mu)E^R(\mu), \qquad
\frac{\partial}{\partial t_j}E^R(\mu)
=m_j(\mu)E^R(\mu),
\]
\[
l_j(\mu)=\mu P_j+[Q,P_j],
\qquad 
m_j(\mu)=-\mu l_j(\mu)
+\frac{\partial Q}{\partial y_j}.
\]
Here we have used the matrix $P_j$ def\/ined in 
(\ref{kojima:def:P}) and
\be
Q=\left(\begin{array}{cc}
\ds -\frac{1}{2\pi}\int_{-\infty}^{\infty}
e^{\tau(s|y_2-y_1,t_2-t_1)}ds \sigma_+&
\ds -\int_{-\infty}^{\infty}
e_1^R(s)e_2^L(s)ds
\vspace{3mm}\\
0&\ds -\frac{1}{2\pi}\int_{-\infty}^{\infty}
e^{\tau(s|y_4-y_3,t_4-t_3)}ds \sigma_+ \end{array}\right),
\label{kojima:def:Q}
\ee
where $\tau(s|y,t)$ is def\/ined in (\ref{kojima:def:tau}).
Def\/ine the integral operator $\widehat{M}$ by
\[
\left(\widehat{M}f\right)(\lambda)
=\int_0^q M(\lambda,\mu)f(\mu)d\mu,
\]
where we have used the kernel def\/ined by
\be
M(\lambda,\mu)=-
\frac{\pi}{2}\frac{E^L(\lambda)E^R(\mu)}{\lambda-\mu}.
\label{kojima:def:M}
\ee
By direct calculations we obtain the following Propositions.

\begin{prop} The kernel $L(\lambda,\mu)$
is the special case of the kernel $M(\lambda,\mu)$
\[
L(\lambda,\mu|t,x_1,x_2)
=M\left(\lambda,\mu 
\left|\begin{array}{cccc}-x_1&x_1&-x_2&x_2\\
0&0&t&t
\end{array}\right.\right).
\]
\label{kojima:prop:deg1}
\end{prop}

Def\/ine the vectors 
$F^L(\lambda)$ and $F^R(\mu)$ by the integral equations
\[
F^L(\lambda)=
\Bigl(
F^L_1(\lambda)~~F^L_2(\lambda)~~
F^L_3(\lambda)~~F^L_4(\lambda)
\Bigr)=
-\left(\left(1-\frac{2}{\pi}\widehat{M}\right)^{-1}
E^L\right)(\lambda),
\]
\[
F^R(\mu)=
\left(
\begin{array}{c}
F^R_1(\mu)
\vspace{1mm}\\
F^R_2(\mu)
\vspace{1mm}\\
F^R_3(\mu)
\vspace{1mm}\\
F^R_4(\mu)
\end{array}\right)=
\left(E^R\left(1-\frac{2}{\pi}\widehat{M}\right)^{-1}\right)
(\mu).
\]
By usual calculation procedure described in \cite{kojima:K.B.I.}, we 
obtain the closed dif\/ferential equations
\be
\frac{\partial}{\partial y_j}F^R(\mu)
=L_j(\mu)F^R(\mu),\qquad
\frac{\partial}{\partial t_j}F^R(\mu)
=M_j(\mu)F^R(\mu),\label{kojima:oLax}
\ee
\[
L_j(\mu)=\mu P_j+[b,P_j],
\qquad M_j(\mu)=-\mu L_j(\mu)
+\frac{\partial b}{\partial y_j}.
\]
Here we have used matrix $b$ def\/ined by
\be
b=B+Q,\label{kojima:def:b}
\ee
where $Q$ is def\/ined in (\ref{kojima:def:Q}) and
the matrix $B$ is def\/ined by
\[
B_{j,k}
=B_{j,k}\left(
\begin{array}{cccc}y_1&y_2&y_3&y_4\\
t_1&t_2&t_3&t_4\end{array}\right)
=
\int_{-q}^q F^R_j(\lambda)
E^L_k(\lambda)d\lambda.
\]
The compatibility condition of the above dif\/ferential equations
(\ref{kojima:oLax})
yields the dif\/ferential equations (\ref{kojima:Lax}).

\begin{prop} A factor 
of correlation functions can be written by 
an element of the matrix $b$
\[
G(x_1+x_2)-
\frac{1}{2\pi}
{\rm Tr}\left(\left(1+\frac{2}{\pi}\widehat{S}\right)
\widehat{Q}\right)=b_{1,4}\left(
\begin{array}{cccc}-x_1&x_1&-x_2&x_2\\
0&0
&t&t\end{array}\right).
\]
\end{prop}

\noindent 
{\sl Proof.} The kernel $U(\lambda,\mu)$ (\ref{kojima:def:U})
is related
to the vectors $E^R(\lambda)$ and $E^L(\mu)$
\[
Q(\lambda,\mu)=-2\pi \left.E^R_1(\lambda) E^L_4(\mu)
\right|_{
y_1=-y_2=-x_1; y_3=-y_4=-x_2;
t_1=t_2=0 ; t_3=t_4=t}.
\]
By using Proposition \ref{kojima:prop:deg1}, we arrive at the following
\[
{\rm Tr}\left(\left(1+\frac{2}{\pi}\widehat{S}\right)
\widehat{U}\right)=-2\pi B_{1,4}\left(
\begin{array}{cccc}-x_1&x_1&-x_2&x_2\\
0&0
&t&t\end{array}\right).
\]

\subsection*{Q.E.D.}

Now we arrive at Theorem \ref{kojima:Th2}.
For f\/inite temperature case $T>0$,
we prepare some functions.
Let us set
\[
\left(M_T f\right)(\lambda)
=\int_{-\infty}^\infty
M(\lambda,\mu)\vartheta(\mu)d\mu,
\]
where $\vartheta(\mu)$ is def\/ined in (\ref{kojima:def:Fermi}) and
$M(\lambda,\mu)$ is def\/ined in (\ref{kojima:def:M}).
Def\/ine the vectors 
$F^L(\lambda)_T$ and $F^R(\mu)_T$ by the integral equations
\[
F^L(\lambda)_T
=
-\left(\left(1-\frac{2}{\pi}\widehat{M}_T\right)^{-1}
E^L\right)(\lambda),
\qquad
F^R(\mu)_T=
\left(E^R\left(1-\frac{2}{\pi}\widehat{M}_T\right)^{-1}\right)
(\mu).
\]
Def\/ine the matrix $b^T$ by
\be
b^T=B^T+Q,\label{kojima:def:bT}
\ee
where $Q$ is def\/ined in (\ref{kojima:def:Q}) and
the matrix $B^T$ is def\/ined by
\[
B_{j,k}^T
=B_{j,k}^T\left(
\begin{array}{cccc}y_1&y_2&y_3&y_4\\
t_1&t_2&t_3&t_4\end{array}\right)
=\int_{-\infty}^\infty F^R_j(\lambda)_T
E^L_k(\lambda)_T d\lambda.
\]
By the similar discussion as temperature $T=0$ case,
we arrive at Theorem \ref{kojima:Cor2}.

\setcounter{equation}{0}

\section{The time-independent Case}
The purpose of this 
section is to derive the Fredholm minor determinant
representations for f\/inite-temperature 
f\/ields correlation functions:
\[
\langle \psi(x_1)\psi^\dagger(x_2)\rangle_{\epsilon,T}.
\]
Our Fredholm minor determinant representations
coincide with the one which has been obtained in~\cite{kojima:K}.
When we take the limit $t \to 0$, the following 
simplif\/ications occur:
\[
\ba{l}
\ds G(x)\to 0, \qquad
L(\lambda,\mu)\to 
\frac{1}{\lambda-\mu}\left(
\sin(x_1(\lambda-\mu))+\sin(x_2(\lambda-\mu))\right),
\vspace{3mm}\\
\ds P(\lambda \vert x_1,x_2)\to e^{-i x_1 \lambda}.
\ea
\]
Therefore we obtain
\[
\ba{l}
\ds \langle \psi(x_1)\psi(x_2) \rangle_{\epsilon,T}
=\left.\frac{1}{2\pi}
\left(\frac{\partial}{\partial \alpha}\right)
\det\left(1-\frac{2}{\pi}\widehat{\tilde{V}}_{\epsilon,T}
-\alpha \widehat{\tilde{W}}^{(x_1,x_2)}_{\epsilon,T}\right)
\right|_{\alpha=0}
\vspace{3mm}\\
\ds \qquad =-\frac{1}{2\pi}
\det\left(1-\frac{2}{\pi}\widehat{\tilde{V}}_{\epsilon,T}\right)
{\rm Tr}\left[\left(1-\frac{2}{\pi}
\widehat{\tilde{V}}_{\epsilon,T}\right)^{-1}
\widehat{\tilde{W}}_{\epsilon,T}^{(x_1,x_2)}\right].
\ea
\]
Here the integral operators are given by
\[
\ds \left(\widehat{\tilde{V}}_{\epsilon,T}f\right)
(\lambda)=\int_0^{\infty}
\tilde{V}_{\epsilon,T}(\lambda,\mu)f(\mu)d\mu,
\qquad  \left(\widehat{\tilde{W}}_{\epsilon,T}^{(\xi,\eta)}
f\right)(\lambda)=\int_0^{\infty}
\tilde{W}_{\epsilon,T}^{(\xi,\eta)}(\lambda,\mu)f(\mu)d\mu,
\]
where the integral kernels are given by
\[
\ba{l}
\ds \tilde{V}_{\epsilon,T}(\lambda,\mu)
=\sqrt{\vartheta(\lambda)}\left[
\frac{1}{\lambda-\mu}\left\{\sin(x_1(\lambda-\mu))+
\sin(x_2(\lambda-\mu))\right\}\right.
\vspace{3mm}\\
\ds \qquad +\left.\epsilon \frac{1}
{\lambda+\mu}\left\{\sin(x_1(\lambda+\mu))+
\sin(x_2(\lambda+\mu))\right\}\right]
\sqrt{\vartheta(\mu)},
\ea
\]
\[
\tilde{W}_{\epsilon,T}^{(\xi,\eta)}
(\lambda,\mu)=\sqrt{\vartheta(\lambda)}
\epsilon(e^{i \xi \lambda}+\epsilon e^{-i \xi \lambda})
(e^{i \eta \mu}+\epsilon e^{-i \eta \mu})\sqrt{\vartheta(\mu)}.
\]
Pay attention to the Fourier transforms:
\[
f(\lambda)=\frac{1}{2 \pi \sqrt{\vartheta(\lambda)}}
\int_{-\infty}^{\infty} d\xi e^{i\lambda \xi}\varphi(\xi),
\qquad
\varphi(\xi)=\int_{-\infty}^{\infty} d\lambda 
\sqrt{\vartheta(\lambda)}
e^{-i\lambda \xi}f(\lambda).
\]
The following identity holds for functions 
$f_{\epsilon}(\epsilon \lambda)=\epsilon f(\epsilon \lambda)$:
\[
\ba{l}
\ds \int_0^{\infty}d\mu f_{\epsilon}(\mu)
\sqrt{\vartheta(\lambda)\vartheta(\mu)}
\left\{\frac{1}{\lambda-\mu}\sin(x(\lambda-\mu))+
\epsilon \frac{1}{\lambda+\mu}\sin(x(\lambda+\mu))\right\}
\vspace{3mm} \\
\ds \qquad =
\frac{1}{2\pi\sqrt{\vartheta(\lambda)}}
\int_{-\infty}^{\infty}d\xi e^{i\lambda \xi}
\left(\int_0^x d\xi' \theta_{\epsilon,T}(\xi',\xi)\right)
\int_{-\infty}^{\infty}d\mu
f_{\epsilon}(\mu)\sqrt{\vartheta(\mu)}e^{-i\xi'\mu},
\ea
\]
where
\[
\theta_{\epsilon,T}(\xi,\eta)
=\int_0^{\infty}\vartheta(\nu)
\left\{\cos((\xi-\eta)\nu)+
\epsilon \cos((\xi+\eta)\nu)\right\}d\nu.
\]
Therefore we arrive at 
\[
\det\left(1-\frac{2}{\pi}\widehat{\tilde{V}}_{\epsilon,T}\right)
=\det\left(1-\frac{2}{\pi}\left(
\hat{\theta}_{\epsilon,T}^{(x_1,x_2)}
\right)\right),
\]
where the integral operator 
$\hat{\theta}_{\epsilon,T}^{(y_1,y_2)}$
is def\/ined by
\[
\left(\hat{\theta}_{\epsilon,T}^{(y_1,y_2)}f\right)(\xi)
=\int_{0}^{\infty}
\left(({\rm E}(y_1-\xi')+{\rm E}(y_2-\xi'))
\theta_{\epsilon,T}(\xi,\xi')\right)f(\xi')d\xi'.
\]
Here ${\rm E}(\xi)$ represents the step function
\[
{\rm E}(\xi)=\left\{
\begin{array}{cc}1, & {\rm for} \quad \xi \geq 0, \\
0, & {\rm for} \quad \xi < 0.
\end{array}\right.
\]
Let us set
\[
\Delta_{\epsilon}(\xi,\eta)=\frac{\det
\left(1-\frac{2}{\pi}\left(
\hat{\theta}_{\epsilon,T}^{(x_1,x_2)}\right)\left|
\begin{array}{c}\eta\\
\xi
\end{array}\right.\right)}
{\det\left(1-\frac{2}{\pi}\left(
\hat{\theta}_{\epsilon,T}^{(x_1,x_2)}
\right)\right)}.
\]
Here we have used the following notation of the $r$-th 
Fredholm minor determinants:
\[
\ba{l}
\ds \det
\left(1-\lambda \widehat{K}_I\left|
\begin{array}{ccc}\xi_1 & \cdots & \xi_r\\
\eta_1 & \cdots &\eta_r
\end{array}\right.\right)
\vspace{3mm}\\
\ds \qquad =
\sum_{n=0}^{\infty}\frac{(-\lambda)^{n+r}}{n!}
\int_I d\lambda_1 \cdots 
\int_I d\lambda_n K_{n+r}\left(\begin{array}
{cccccc}\xi_1 & \cdots & \xi_r 
~\lambda_1 &\cdots &\lambda_n\\
\eta_1 & \cdots &\eta_r
~\lambda_1 &\cdots &\lambda_n
\end{array}\right),
\ea
\]
where we have used
\[
K_{m}\left(\begin{array}{ccc}\xi_1 & \cdots & \xi_m \\
\eta_1 & \cdots &\eta_m
\end{array}\right)
=\det_{1 \leq j,k \leq m}
\left(K(\xi_j,\eta_k)\right).
\]
The integral operator $\widehat{K}_I$ is def\/ined by using
the integral kernel $K(\lambda,\mu)$ and the 
integral interval $I$:
\[
(\widehat{K}_I f)(\lambda)=\int_I K(\lambda,\mu)f(\mu) d\mu.
\]

From the above def\/inition of the Fredholm minor determinants,
the function $\Delta_{\epsilon}(\xi,\eta)$ satisf\/ies 
the integral equation:
\[
\Delta_{\epsilon}(\xi,\eta)-\frac{2}{\pi}
\int_{-x_1}^{x_2}\theta_{\epsilon,T}(\xi,\xi')
\Delta_{\epsilon}(\xi',\eta)d\xi'=
-\frac{2}{\pi}\theta_{\epsilon,T}(\xi,\eta).
\]
Let us take the Fourier transforms of this integral equation
\[
\ba{l}
\ds \frac{1}{2\pi \sqrt{\vartheta(\lambda)}}
\int_{-\infty}^{\infty}d\xi e^{i\lambda \xi}
\Delta_{\epsilon}(\xi,\eta)
-\frac{2}{\pi}\int_{0}^{\infty}d\mu \tilde{V}_{\epsilon,T}
(\lambda,\mu)
\frac{1}{2\pi \sqrt{\vartheta(\mu)}}
\int_{-\infty}^{\infty}d\xi'e^{i\mu \xi'}
\Delta_{\epsilon}(\xi',\eta)
\vspace{3mm}\\
\ds \qquad =-\frac{1}{\pi}\sqrt{\vartheta(\lambda)}(e^{i\lambda \eta}+
\epsilon e^{-i\lambda \eta}).
\ea
\]
Therefore we obtain
\[
\Delta_{\epsilon}(\xi,\eta)=
-\frac{1}{\pi}{\rm Tr}\left[
\left(1-\frac{2}{\pi}\widehat{\tilde{V}}_{\epsilon,T}\right)^{-1}
\widehat{\tilde{W}}_{\epsilon,T}^{(\xi,\eta)}\right].
\]
Now we arrive at Theorem \ref{kojima:Th3}.
In much the same way as with f\/inite temperature case
$T>0$ we arrive at Theorem \ref{kojima:Cor3}.

\subsection*{Acknowledgements}
I wish to thank to Professor Miwa for his encouragements.
This work is partly supported by the grant from the Research
Institute of Science and Technology, Nihon University.

\label{kojima-lp}

\end{document}